------------------------------INSTRUCTIONS-------------------------------
There are 3 files:

	(1) the latex source file;
	(2) the ccsr.sty file; 
	(3) the figures.uu file.

Save the first file as, say, paper.tex, and save the second file as ccsr.sty.
Keep the two in the same directory. Then run:

 latex paper

Do it twice. You get a paper.dvi file which can be printed in whatrever way
according to your local computer environment.

Save the third file as figures.uu, then run

 csh figures.uu 

which generates 17 postscript files (fig1a.ps....fig5c.ps). Print each
of them to get a figure.

If you have any problems, please contact me (wli@cshl.org) and I can
send you the postrscript files directly.
------------------------------end of instruction-------------------------------

----------------------------cut (file 1)-----------------------------------
\documentstyle{ccsr}

\renewcommand{\baselinestretch}{1.18} 
\begin{document}
 
\title{Understanding Long-range Correlations in DNA Sequences}
\author{
 Wentian Li, Thomas G. Marr\\
 {\small \sl Cold Spring Harbor Laboratory, P.O. Box 100,
 Cold Spring Harbor, NY 11724, USA } \\
 {\small \sl and}\\
 Kunihiko Kaneko \\
 {\small \sl Department of Pure and Applied Sciences}\\
 {\small \sl  University of Tokyo, Komaba, Meguro, Tokyo 153, Japan}
}
\date{(February 17, 1994. to be published in {\sl Physica D})}
\maketitle    
\markboth{\sl Li/Marr/Kaneko}{\sl Long-range Corr}

\begin{abstract}
In this paper, we review the literature on statistical
long-range correlation in DNA sequences. 
We examine the current evidence for these
correlations, and conclude that a mixture of many 
length scales (including some relatively long ones) in DNA sequences 
is responsible for the observed $1/f$-like spectral component. 
We note the complexity of the correlation structure in 
DNA sequences. The observed complexity often makes it hard, or
impossible, to decompose the sequence into a few statistically
stationary regions.  We suggest that, based on the complexity
of DNA sequences, a fruitful approach to understand 
long-range correlation is to model duplication, and other
rearrangement processes, in DNA sequences. One model, called
``expansion-modification system", contains only point
duplication and point mutation. Though simplistic, this model is
able to generate sequences with $1/f$ spectra. We emphasize the importance
of DNA duplication in its contribution to the observed long-range
correlation in DNA sequences.
\end{abstract}

\newpage

\section{Introduction}

During the past year, the study of three groups indicated the
existence of statistical long-range base-base correlation in 
DNA sequences \cite{wli-kuni,wli-dna,peng,voss}. What is
surprising is not just the existence of long-range correlation,
but the particular form of the correlation structure ---
the $1/f$-like spectral component \cite{wli-kuni,voss}. 
The $1/f$-like spectral component observed in DNA sequences
is not the sole spectral component (the white noise component
is also a major contributor to the spectrum), so it somehow
differs from other $1/f$ noise observed in nature (see \cite{press}
for an early review and \cite{musha} for some recent work on
$1/f$ noise) including music \cite{voss75}. Nevertheless, 
there are many different correlation length scales in DNA sequences. 
These different length scales need not be continuously distributed, so one
does not need to introduce the concept of ``fractal".  Also, the
longest length scale is usually bounded, but one can still observe
the ``long-range" correlation as long as the longest length
scale is not an order of magnitude smaller than the sequence length.

Since the publication of the first round of papers, there appear to be
some misunderstandings of the finding. We list a few here
to clarify the situation.  Some of them will be discussed
more in the later sections.
\begin{itemize}

\item
Statistical correlation is not equivalent to a ``cause-and-effect"
correlation. A long-range statistical correlation in DNA sequences means
the base appearance or base density tends to co-vary at regions separated
by a long distance. It is a completely different concept when
one says that one region of the DNA sequence has a biological
influence on another that is far apart.

\item
The current studies only clearly show the existence of a statistical
correlation up to the scale of 1 -- 10  kb 
(1 kb = 1000 bases). The correlation structure of DNA sequences
at even larger scales has not yet been thoroughly studied.

\item
These studies point out not only the existence of long-range
correlation, but much more interestingly, the existence of
a particular form of correlation structure in DNA sequences.
The non-trivial nature or the complexity of the correlation structure is
illustrated by the $1/f$-like spectral component, an indication
of being multi-length-scaled. Trivial or simple  long-range correlation
can be easily generated by alternating large regions, of more or less
the same length, with totally different base compositions.  But
such single-length-scaled sequences will not exhibit $1/f$-like
spectral component.

\end{itemize}

We start in section 2 by reviewing some of the measurements used
to study the correlation structure of DNA sequences, then we
graphically illustrate the main theme that there are fluctuations
of base densities at many different length scales (section 2.1).
Later, we review the current knowledge about the correlation
structure of DNA sequences (section 2.2). We try to reconcile
different results, and try to provide an explanation of 
detailed observations.

We review three ``controversial" topics in section 2.3. The first is
on how good the scale invariant property of DNA sequences
is. The second is on whether the DNA sequences with
long-range correlation can be decomposed into sub-regions
of different base densities, each with sub-regions of 
white noise with no correlations. The third is on
whether coding sequences have a common correlation structure
that is different from that derived from non-coding sequences.

The second topic mentioned above is a question on whether
DNA sequences are simple or ``complex". If the sequence is
basically simple, its correlation structure should be easily
characterized by a decomposition of the sequence into
spatially separable simple sub-sequences (e.g., each sub-sequence
is  white noise).  On the other hand, if a sequence is complex, then
there are many length scales, each of which acquires
its contribution from throughout the sequence. When a spatial decomposition
is impossible, we can decompose the sequence in the frequency
domain; which is simply the classical spectral analysis. We show
that DNA sequences are usually complex.

For objects as complex as DNA sequences, we might adopt 
a different paradigm --- to simulate the process that generates
the sequence.  This is what we call the ``dynamical origin" of the correlation
structure of DNA sequences,  which is the topic discussed
in section 3. From the study of nonlinear dynamical systems,
we know that complexity stored in an object can
be created by a relatively simple process. For example,
imagine a dynamical process that doubles the whole sequence
(plus a random mutation afterward). Although the doubling
process is extremely simple, it actually generates
sequences that are multi-length-scaled.

In section 3.1, we review a model (expansion-modification systems \cite{wli89,wli91})
that is also seemingly simple, but is capable of
generating sequences with a fair amount of complexity:
in this case, the $1/f$ spectrum. In this model, the only processes
involved  are a point duplication and a point mutation, each
occurring with a fixed probability. Although this model
is simplistic in describing changes in DNA sequences,
both duplication and point mutation are common events for DNA sequences.
In fact, the original reason for our search for $1/f$ spectra in 
DNA sequences came from the study of the expansion-modification
systems \cite{wli-kuni-2}.  Since both processes can create new
patterns in the sequence, these are potentially important
for evolution. This topic is discussed in section 3.2.

In summary, we use this opportunity to review what we currently 
know about the statistical long-range correlation in DNA sequences.
There are many issues one can question: whether long-range correlation is
a common feature of DNA sequences; whether the correlation
structure of DNA sequences at 1 Mb (1 Mb= $10^6$ bases)
is similar to that at 1 kb;  whether the correlation structure
has any relevance to our understanding of biological function;
whether the statistical structure of 1-dimensional
DNA sequences can reveal features of the 3-dimensional structure 
of the chromosome, and so on. We do not have definitive
answers to these questions yet. With more and more long
stretchs of DNA sequences become available, it is an exciting
time for those who are interested in exploring and understanding
the large-scale  statistical structure of DNA sequences.

\section{Statistical Correlation Structures of DNA Sequences}

Correlation structure of a length-$N$ DNA sequence is the correlations
between two nucleotides of any distance $d$ ($<N$).  The intention here
is to characterize the correlation structure of a particular DNA
sequence with a finite length, rather than to generalize the result
to infinitely long sequences, nor to a sub-region of the sequence.
So we are not concern about the concept of ``statistical stationarity". 
It is similar to the case when one says that a sample density of G
of a particular DNA sequence is 0.21: it does not necessarily imply
that the density of G of other sequences is also 0.21, nor does it
claim that any sub-region of length $n <N$ has the same density of 0.21.

Correlation structure of a finite sequence can be studied
by several sample measures of correlation. We list these in
three categories:

\vspace{0.4in}
{\bf Direct measures of correlation}

We first define a sample ``mutual information function": if
the indices $\alpha$ and $\beta$ run through the four
nucleotides (G,C,T,A), we define (the cyclic boundary is
used)\footnote{
An alternative sampling method is to not use the cyclic
boundary condition, so
$n_{\alpha \beta}(d) \equiv \sum_{i=1}^{N-d} 1[ x_i=\alpha 
\hspace{0.1in} \mbox{and} \hspace{0.1in} x_{i+d}=\beta]$.
Because the number of point sampled is less than $N$,
one has to introduce some correction factor to compensate
this fact.
}
\begin{equation}
\label{eq:ab}
n_{\alpha \beta}(d) \equiv \sum_{i=1}^{N} 
1[ x_i=\alpha \hspace{0.1in}  \mbox{and} \hspace{0.1in}  x_{i+d}=\beta]
\end{equation}
as the total number of counts of a particular joint nucleotide-pair
type ($\alpha$, $\beta$) with the two separated by a fixed distance $d$, and
\begin{equation}
\label{eq:a}
n_\alpha \equiv \sum_{i=1}^N 1[ x_i = \alpha]
\end{equation}
as the number of counts of the nucleotide type $\alpha$. Note
that $n_{\alpha\alpha}(0) = n_\alpha$.  

The sample mutual information function can be defined as (the
notation $\hat{}$ is used to indicate an estimate)\footnote{
Other alternatives include: (1) to use $\log_{10}$ instead of $\log_e$;
and (2) to normalize the function by its value at $d=0$, i.e.,
$\widehat{m}(d) = \widehat{M(d)}/\widehat{M(0)}$.
}:
\begin{eqnarray}
\widehat{M(d)} & = &
 \frac{1}{N} \sum_{\alpha=(G,C,T,A)} \sum_{\beta=(G,C,T,A)}
n_{\alpha\beta}(d) \log \left( \frac{N n_{\alpha\beta}(d)}{ n_\alpha n_\beta} \right)
\nonumber \\
&=& log N + \frac{1}{N}  \sum_{\alpha=(G,C,T,A)} \sum_{\beta=(G,C,T,A)}
n_{\alpha\beta}(d) \log \left( \frac{n_{\alpha\beta}(d)}{ n_\alpha n_\beta} \right).
\end{eqnarray}

This sample mutual information function is an estimator of
the mutual information function:
\begin{equation}
M (d) \equiv \sum_\alpha \sum_\beta P_{\alpha\beta}(d) \log
\left( \frac{ P_{\alpha\beta}(d)}{ P_\alpha P_\beta}
\right),
\end{equation}
where $P_{\alpha \beta}(d)$ is the joint probability of ($\alpha$, $\beta$)
symbol pair, and $P_\alpha$ is the density of symbol $\alpha$.
This quantity was first introduced in information theory \cite{shannon},
and recently is applied to the study of chaotic nonlinear dynamics
\cite{mutu-shaw,fraser,mutu-kuni,herzel1}, symbolic sequence analysis
\cite{herzel2,mutu-wli}, learning features from experiments
\cite{fred}, nonlinear prediction \cite{meyer},  improving neural network
performance \cite{lapedes},
identifying active sites in AIDS virus sequence \cite{bette}, for example.
Mutual information is now considered a standard  measure of correlation
(see page 634 of \cite{recipe}).  The famous ``PAM" (Accepted Point Mutation)
matrix element used to measure the point mutation rate from one amino acid
to another is of the form of log-likelihood \cite{dayhoff2}, and the
average of all these matrix elements is exactly a mutual
information \cite{mutu-altschul}. See also a textbook introduction
in \cite{sakamoto}

One may also use the traditional covariance or autocorrelation
function to calculate the same-symbol correlation (note that no cross
correlation between different symbols is included). Here is a
sample covariance function:
\begin{equation}
\label{eq:est-cov}
\widehat{cov(d)} = \sum_{\alpha =(G,C,T,A)} 
\left( \frac{ n_{\alpha\alpha}(d)}{N} - \frac{ n_\alpha^2}{N^2}
\right),
\end{equation}
which is an estimator of the covariance function:
\begin{equation}
\label{eq:cov}
cov(d) \equiv  \sum_{\alpha =(G,C,T,A)}
\left(
P_{\alpha\alpha}(d) -P_\alpha^2
\right).
\end{equation}
And here is a sample autocorrelation function --- the sample covariance
function normalized by its value at $d=0$:
\begin{equation}
\widehat{\Gamma(d)} =
\sum_{\alpha =(G,C,T,A)} 
\left(  \frac{ n_{\alpha\alpha}(d) - n_\alpha^2 /N}
{ n_\alpha - n_\alpha^2/N}
\right),
\end{equation}
which is an estimator of the autocorrelation function:
\begin{equation}
\Gamma(d) \equiv  \frac{\sum_{\alpha =(G,C,T,A)} \left( P_{\alpha\alpha}(d)-P_\alpha^2 \right)}
{\sum_{\alpha =(G,C,T,A)} \left( P_{\alpha}-P_\alpha^2 \right)}.
\end{equation}
Again, there are several alternatives, either to the sampling method
(e.g., use of the cyclic boundary condition), or to the definition
of covariance function for DNA sequences (e.g., take the absolute
value for each term in Eq.(\ref{eq:cov}) before adding these together).

\vspace{0.4in}
{\bf Frequency domain characterization}

Due to various reasons (see, e.g., pages 7-8 of \cite{percival}, 
including being easier to interpret the result),
power spectrum, instead of autocorrelation function, is often used
to characterize the statistical structure of a sequence. Here is
a sample measure of power spectrum of DNA sequence:
\begin{equation}
\widehat{S_k} = \frac{1}{N^2} \sum_{\alpha=(G,C,T,A)}
\left|
\sum_{j=1}^N 1 [x_j=\alpha] e^{ -i 2 \pi \frac{k}{N} j }
\right|^2.
\end{equation}
The frequency is $f=k/N$ ($k=1,2, \dots N/2$). There are
also other definitions of power spectrum for multi-symbol sequences
(e.g., \cite{silver}). We usually call
a sequence ``$1/f^\alpha$ spectrum" if its power spectrum
behaves as an inverse power-law with the exponent
$\alpha$: $S_k \sim 1/k^\alpha$. And say a sequence
contains a ``$1/f^\alpha$ spectral component" if the
spectrum is mainly a ``$1/f^\alpha$ spectrum" plus a
white noise: $S_k \sim S_{N/2} + 1/f^\alpha$
\cite{traffic,voss76}. $1/f^\alpha$
spectrum is non-trivial only if the exponent of the
power-law function $\alpha$ is close to 1.  If $\alpha
\approx 0$ we have white noise, and if $\alpha
\approx 2$, we have a random walk sequence, which is
also rather easy to generate (e.g., \cite{wli-dow}).

\vspace{0.4in}
{\bf Cumulative variables}

The approach of using cumulative variables to study the correlation
structure of DNA sequences can suppress fluctuations in $cov(d)$
or $\Gamma(d)$ \cite{peng}, but it measures essentially the same
thing. Following Ref.\cite{peng}, if we define the purine-pyrimidine
sequence as (purine =(A,G), pyrimidine =(C,T)):
\begin{equation}
b_i \equiv \left\{
\begin{array}{ll}
1 & \mbox{if $x_i$ = $(A,G)$} \\
0 & \mbox{if $x_i$ = $(C,T)$} \\
\end{array}
\right.
\end{equation}
and the cumulative variable $y_k(l)$ is the sum of $b_i$
from $i=k$ to $i=k+l-1$:
\begin{equation}
y_k(l) \equiv \sum_{i=k}^{k+l-1} b_i,
\end{equation}
then the variance of $ y_k(l) $ series ($k=1, 2, \cdots, N-l+1$) ---
$var_y(l)$ (subscript $y$ indicates the $y$ series)--- can be estimated  by
\begin{equation}
\widehat{var_y(l)} = \frac{1}{N} \sum_{k=1}^{N-l+1} y_k(l)^2
- \frac{1}{N^2} \left( \sum_{k=1}^{N-l+1} y_k(l) \right)^2.
\end{equation}
It can be shown that $var_y(l)$ is a sum of covariance
functions from the original purine-pyrimidine sequence \cite{peng,karlin}
(subscripts $b$ indicates the original binary series):
\begin{equation}
\label{eq:cumu}
var_y(l) = l \cdot cov_b(0) + 2(l-1) \cdot cov_b(1) + 2(l-2) \cdot cov_b(2) \cdots
+2 \cdot cov_b(l-1).
\end{equation}
If $\{ b_i \}$ is a white noise, all $cov_b(d)$'s are zero except $cov_b(0)$.
So $var_y(l)$ is a linear function of $l$. If $\{ b_i \}$ is generated
by a first-order  Markov chain with only 1-step correlations, $cov_b(d)$
decreases with $d$ exponentially. In this case, $var_y(l)$ is almost
linearly proportional with $l$, plus an exponential correction. If
$cov_b(l)$ decays slower than exponential functions, $var_y(l)$
may not change with $l$ linearly. In particular, a power-law function
in $cov_b(d)$ leads to a power-law function in $var_y(l)$.

We notice that
\begin{itemize}
\item
One might first calculate the covariance function $cov_b(d)$ or autocorrelation function
$\Gamma_b(d)$ from the original sequence, then derive $var_y(l)$ according
to Eq.(\ref{eq:cumu}). In other
words, the essence of using the cumulative variable $y_k(l)$ is to
sum over autocorrelation function at different distances so that the
fluctuation in the autocorrelation function is reduced.

\item
If $cov_b(d)$ oscillates between positive and negative values, the
cumulative effect on $var_y(l)$ may be small. One might wonder whether
it can happen that even when the amplitude of the oscillation decays slower
than exponential, the $var_y(l)$ is still a linear function of $l$. 

\item
A deviation from the linear behavior in $var_y(l)$ for a set of
finite $l$ values only requires a slow decay of the $cov_y(d)$
or $\Gamma_b(d)$ within this particular range of $l$. One should be cautious
when making claims about power-law functions or fractal functions, that mainly
make sense in the $l \rightarrow \infty$ limit. For finite sequences, 
we can only take the $l \rightarrow N$ limit, but this limit
is complicated by the finite-size effects (see, e.g., \cite{peng-finite}).

\end{itemize}

Any one of the three approaches (direct calculation, frequency domain
characterization, and cumulative sequence) can be used to analyze
the correlation structure of DNA sequences. There are other
approaches as well. In the following, we will first visualize
the base composition fluctuation in some long DNA sequences,
then review the study of correlation structure that has appeared
in recent publications, and discuss some issues currently
under debate.

\subsection{Large Scale Fluctuations in Base Composition}

One of the easiest ways to  obtain information from a DNA sequence
is to ``view" it.  Good visualization can help us develop intuition about
the sequence. Since the study of correlation structure is
essentially the study of fluctuation of base densities at different 
length scales,  we view the base densities within a moving window
of different sizes ($W$). Besides the window size, another
parameter is the shifting distances from one window to another ($D$).

To simplify the work, we plot two types of densities
instead of four base densities. The first is the density for
G or C (``G+C content") within the $j$th moving window:
\begin{equation}
\rho_{G+C}(j) = \frac{1}{W} \sum_{i=(j-1)D+1 }^{(j-1)D+W} 1 [ \mbox{if $x_i = (G, C)$} ]
\hspace{0.3in} \mbox{$j=1, 2, \cdots$}.
\end{equation}
If the moving distance $D$ is equal to the window size $W$, windows
are non-overlapping; if $D<W$, windows are overlapping. Similarly
we can calculate the purine density:
\begin{equation}
\rho_{purine}(j) = \frac{1}{W} \sum_{i=(j-1)D+1 }^{(j-1)D+W} 1 [ \mbox{if $x_i = (A, G)$} ]
\hspace{0.3in} \mbox{$j=1, 2, \cdots$}.
\end{equation}

Figure 1 shows $\rho_{G+C}$ and $\rho_{purine}$ for the complete
DNA sequence of {\sl Saccharomyces cerevisiae} (also called budding yeast)
chromosome 3 (GenBank locus name: SCCHRIII; accession number: X59720) \cite{oliver}
with various window sizes $W$ and/or moving distances $D$.
The sequence length is $N=$ 315,338 bases. The average density of $G+C$,
0.3855, and the average density of purine, 0.4998, are drawn
by the horizontal lines.

Figure 1(a) shows the plot for $W=1000$ and $D=100$, a case with a high
degree of overlaps between neighboring windows. It represents a fine
scaled plot of base composition fluctuation because both $W$ and $D$ are
much smaller than the sequence length $N$ ($ W \approx 0.003 N$
and $D \approx 0.0003 N$). Figure 1(b) is
a case for non-overlapping windows: $W=D=1000$. Figure 1(b) is very
similar to Figure 1(a).

In Figure 1(c) and (d), the window size is increased to $W=10000$ and $W=20000$
while $D$ is kept at 1000. At $W = 20000$, it can be seen that
$G+C$ content is smaller at both two telomeres and the centromere while
larger in between. This alteration of $G+C$ rich and poor regions
is reminiscent of the ``isochore" in frog, chicken, mouse, and human
genomes studied in Bernardi's group \cite{isochore}, though the
size of the two $G+C$ rich regions in budding yeast chromosome 3
( $\sim$ 60  kb) is smaller than that of isochores ( $\sim$ several
hundred kb).\footnote{
An interesting observation of the budding yeast chromosome 3
sequence is that only coding region is responsible for the
$G+C$ rich/poor variations \cite{sharp}. The large-scale
variation of $G+C$ content along the sequence is not observed
in non-coding segments \cite{sharp}.
}

Figure 1(e) goes to an extreme with $W=100,000$ and $D=2000$. Since
the sequence length is around 300,000, the non-overlapping
window version of this plot (i.e., if $D=100,000$) will only show three points.
Roughly speaking, for $G+C$ content, two points are above average, and one
is below average.  For purine density, one point is above average, and one
is below average.

Figure 2 shows the similar density plots for the complete sequence of
a human cytomegalovirus (strain AD169, with the GenBank locus name: HEHCMVCG,
and  accession number X17403) \cite{ad169}. The sequence length is $N=229,354$.
Window sizes are $W=$ 1000, 5000, and 10000, respectively, and moving
distance is kept at $D=1000$. The statistical structure of this
sequence is studied in \cite{voss}.

Figure 3 shows the density plots for the complete sequence of human blood 
coagulation factor VII gene (GenBank locus name: HUMCFVII, and
accession number: J02933) \cite{blood}. The sequence length is $N=12,850$.
Window sizes are $W=$ 30, 100, and 1000, respectively, while moving
distance is kept at $D=30$. The statistical structure of this
sequence is studied in \cite{wli-kuni,wli-dna}.

Finally, Figure 4 shows the density plots for the complete
sequence of bacteriophage lambda (GenBank locus name: LAMCG;
accession number: J02459) \cite{lambda}. The sequence length is $N=48,502$.
Window sizes are $W$=100, 2000, and 10000, respectively; and the moving distance
is $D=$100.  References that study the statistical structure
of this sequence  include \cite{peng}, \cite{karlin}, \cite{peng2}
and \cite{larhammar}.

>From these figures, we see that typically there are fluctuations of base
composition (as represented by the $G+C$ content and purine
density) throughout the sequences. The pattern of these fluctuations
can be very complex: a $G+C$ rich region can contain many small
sub-regions that are $G+C$ poor, and vice versa.  A lack of this
feature usually indicates a lack of many length scales.  For example,
the plots in Figure 4 at window sizes $W$=100, 2000, and 10000 exhibit
very similar features, and the $G+C$ rich region on the left
does not have any sub-regions with a below-average $G+C$ content.
As compared with Figures 1-3, this sequence is relatively simple.
In the next sub-section, we review published statistical studies
that reveal this multi-scale phenomenon in DNA sequences.

\subsection{Published results of statistical correlation structure
of DNA sequences}

In Ref.\cite{wli-dna}, the mutual information function of five
coding sequences and five non-coding sequences from human DNA
is calculated. The sequence length ranges from around 2 kb to
16 kb. It is observed that correlation structure as measured
by the mutual information function can differ from one sequence
to another. For these sequences (ten in all), the correlation
decays to a negligible value at less than 10 bases for coding
sequences, and at slightly more than 10 bases for non-coding
sequences, with the exception of one coding sequence and
one non-coding sequence.

That exception in non-coding sequence has the longest correlation
length among the 10 cases, which is the blood coagulation factor VII
gene whose base composition fluctuation is plotted in Figure 3. We
examine the correlation structure of the sequence in more detail in
Ref.\cite{wli-kuni}, using the complete sequence instead of only
the non-coding region (though 76\% of the sequence is non-coding). 
In particular, it is observed that the power spectrum of the sequence
has a $1/f^\alpha$ ($\alpha \approx 1$) spectral component,
with $\alpha \approx 0.84$, $0.57$
and $0.53$ when the first, the middle, and the last $2^{13}=8192$
bases are taken for the spectral analysis (the sequence length is
$N$=12850). This was the first example that the $1/f^\alpha$ spectral
component in DNA sequence was shown.

In a recent study \cite{borsnik}, it is questioned whether a Lorenzian
spectrum of the form (both a continuous and a discrete version):
\begin{equation}
\label{eq:loren}
S(f) = \frac{ 2 d_0}{ 1 + (d_0 2 \pi f)^2 } \hspace{0.2in} \mbox{or} \hspace{0.2in}
S_k = \frac{2 d_0}{ 1 + ( d_0 2 \pi \frac{k}{N} )^2}
\end{equation}
is ``mistaken" as a $1/f^\alpha$ spectrum component. In \cite{borsnik},
the average  spectrum for flanking sequences (those
located before or after a gene)
is presented (the DNA sequence is converted to a binary sequence, and
it seems that $2^{12}=4096$ bases are taken in calculating
each spectrum), and it is somewhat similar to the result in \cite{wli-kuni}
that the power raises at lower frequencies. An attempt is made to
fit the spectrum by both a Lorenzian spectrum and a $1/f^\alpha$
spectrum. The authors of \cite{borsnik} believe
that the spectrum is essentially a Lorenzian spectrum,
nevertheless it can be fit by a $1/f^\alpha$ spectrum.
But we note that the
$1/f^\alpha$ fit is good for only one decade in the frequency range, 
which is narrower than a typical $1/f$ spectrum. We also note that the very
first point (the power at the lowest frequency) is off the fitting line 
(see Figure 2 of \cite{borsnik}).

We have the following comments:
\begin{itemize}

\item
One of the earliest explanations of $1/f$ noise is that there
are many time scales, with each contributing a Lorenzian spectrum \cite{ziel}.
Since in our case the spectrum is not all $1/f$ --- it also has
a white noise component  --- the number of length scales
required to recreate the spectrum is less.  The issue here is
whether in DNA sequences there is only {\sl one} length scale (that is the case for
a Lorenzian spectrum) or many.

\item
If we superimpose a few Lorenzian spectra, the look of the
overall spectrum depends on how large the longest
length scale is relative to the sequence length.
Figure 5 (a) shows the spectra as used in \cite{borsnik}
(similar to the Lorenzian spectrum, but derived from the series
summation instead of the integration of an exponential function):
\begin{equation}
\label{eq:loren2}
S_k = 1 + \sum_{d_0} a (d_0) \frac{ cos (2 \pi k/N) - e^{-1/d_0}}
{ (e^{1/d_0}+e^{-1/d_0})/2 - cos(2 \pi k/N) } 
\end{equation}
for the three cases for a sequence length of
$N=50,000$: (1) single length scale of $d_0=200$ with
the coefficient $a(d_0=200)=0.03$; 
(2) two length scales:  $a(d_0=200)=0.03$ and $a(d_0=500)=0.01$;
and (3) three length scales with the longest one being
$d_0=2000$: $a(d_0=200)=0.03$, $a(d_0=500)=0.01$
and $a(d_0=2000)=0.007$. 

As expected, adding spectral components
with longer length scales, such as case (3), raises the
spectrum at lower frequencies. But if we repeat the same plot
with a longer sequence length, such as $N=500,000$, even $d_0=2000$ 
in case (3) is not good enough to prevent the flattening out of the spectrum
at lower frequencies (see Figure 5(b)). 

\item
Whether there is a sudden drop or not in the covariance function from
$d=0$ to $d=1$ largely determines whether the high frequency
spectrum is flat or $1/f^2$.  Figure 5(c) shows the spectra similar
to Figure 5(a) except that the normalization condition is
used: $\sum_{d_0} a(d_0)=1$. This condition allows
the exponential autocorrelation function to continuously  extend to
$d=0$. The parameter settings in Figure 5(b) are:
(1) $a(d_0=200)=1$; (2) $a(d_0=200)=0.75$, $a(d_0=500)=0.25$;
and (3)$a(d_0=200)=0.03/(0.03+0.01+0.007)=0.6383$,
$a(d_0=500)=0.2128$, $a(d_0=2000)=0.1489$. One clearly sees the
$1/f^2$ tail replace the flat spectrum at high frequencies.
In this situation, it is unlikely that any part of the spectrum would be mistaken
as a $1/f$ spectrum.

\end{itemize}

The similar $1/f^\alpha$ spectral component is also observed
in a DNA sequence with much longer length: the human cytomegalovirus
\cite{voss}, whose base composition fluctuation is plotted
in Figure 2. The sequence is cut into 3 or 4 pieces, and $2^{16}=65536$ bases
from each piece are used for a spectrum calculation. It should be
noted that correlation structure at length scales longer than
65 kb is not studied in \cite{voss} even though the sequence length
$N$ is approximately 229 kb.

Another interesting study in \cite{voss} is the average
spectrum of {\sl all} sequences in GenBank with $2^{11}=2048$
base in each spectrum calculation. As such, this study reflects
only the average correlation structure up to the length scale of 2 kb.
The exponent $\alpha$ in the $1/f^\alpha$ spectral component
is amazingly close to 1, highly reminiscent of the similar
spectral analysis of music time series \cite{voss75}. A reason why
a much better quality $1/f$
spectral component is observed in this case is that there are
many more different length scales mixed into one statistic, 
and a good $1/f$ spectrum requires many different length scales.

The base-base correlations in a 300 kb long sequence --- the complete
sequence of chromosome 3 of budding yeast, whose base composition
fluctuation is plotted in Figure 1 --- are calculated in \cite{mzhang}
for all possible types of base-pair:
\begin{equation}
\widehat{ I_{\alpha\beta}(d)} = N \frac{n_{\alpha\beta}(d)}{n_\alpha n_\beta}.
\end{equation}
If $\widehat{I_{\alpha\beta}(d)}$ differs from 1, it indicates an
existence of correlation for that particular type of base-pair
at that distance $d$.  It is observed that
$\widehat{I_{\alpha\alpha}(d)}$'s approach 1 from above, while
$\widehat{I_{\alpha\beta}(d)}$'s ($\beta \ne \alpha$) approach 1
usually from below, but sometimes alternatingly from above and below
\cite{mzhang}. Also, $\widehat{I_{\alpha\alpha}(d)}$'s reaches 1
at longer distances than $\widehat{I_{\alpha\beta}(d)}$'s, i.e.,
the longest length scale for yeast chromosome 3 is decided
by the base-pairs of the same type, which is about 1 kb \cite{mzhang}.

\subsection{Review of some controversial topics}

Another line of approach to the study of correlation structure
in DNA sequences first appeared in \cite{peng}. Although it claims less
concerning the correlation structure than  \cite{wli-kuni} and \cite{voss}
(the latter emphasizes the particular correlation structure: the $1/f^\alpha$
spectral component), this paper has generated much more controversy.
In this sub-section, we will focus on reviewing three controversial topics
related to this paper: (1) how good is the scale invariance in the correlation
structure; (2) whether the observed non-white-noise feature in DNA sequences
is purely a consequence of the sequence being composed of sub-regions with
different base densities, whereas each sub-region is a white noise;
(3) to what degree coding and non-coding sequences
have different statistical correlation structures (this topic was actually
suggested earlier in \cite{wli-dna}).

\vspace{0.4in}
{\bf On scale invariance}:

 As we have seen in Eq.(\ref{eq:cumu}),
function $var_y(l)$ changes with $l$ much more smoothly than
function $cov_b(d)$ changes with distance $d$. 
It seems hard to observe $cov_b(d)$ as a power-law
function of $d$ in DNA sequences, but much easier to observe $var_y(l)$ as
a power-law function of $l$, at least up to some upper limit.

Several publications show that the power-law behavior of
$var_y(l)$ breaks down at larger values of $l$'s \cite{prabhu,chat,karlin}.
It is quite natural to assume a mixture of a few length scales
instead of a continuous distribution of length scales as
usually the case for scale-invariant systems. Most important, if there
is an upper limit for the largest length scale, the assumption
of scale invariance is no longer valid beyond that length
scale, and one may not be able to reliably  calculate 
the scaling exponent. See also \cite{munson} for a $var_y(l)$
versus $l$ plot from the budding yeast chromosome 3 sequence.

\vspace{0.4in}
{\bf On whether a complex sequence is decomposable to sub-regions
with simple correlation structures}: 

Based on the observation
of mainly one sequence (bacteriophage lambda, whose base composition
fluctuation is plotted in Figure 4), it is claimed that
the long-range correlation observed in  \cite{peng} can be
fully accounted for by the difference of base compositions
in sub-regions (might also be called ``patches" or ``domains"), such as the
difference of G+C content \cite{karlin}. There are several
problems in this claim. Here are some comments:
\begin{itemize}
\item
Bacteriophage $\lambda$ sequence is not listed as an example of long-range
correlation in \cite{peng}. As seen in both \cite{karlin}
and \cite{peng2}, the $\log ( var_y(l))$ is not a linear function
of $\log(l)$, so both the scaling exponent of 0.61
(it is half of the slope of $\log(var_y(l))$ versus $\log(l)$)
in \cite{larhammar} and 0.53 in \cite{peng} did not
present the whole story. A better example would be a sequence
whose $\log(var_y(l))$ versus $\log(l)$ plot {\sl is} linear but with
a slope different from 1. It is not clear how \cite{karlin}
can use this sequence to illustrate their argument.

\item
As can be seen in Figure 4, the difference of G+C content
between the left half of the sequence and the right half
is clearly visible.  So it is an easy task to remove the effect
of G+C rich/poor regions by using a different average value
of G+C content for each region. It is similar to the
traditional approach to remove the trend from a non-stationary
time series to make it stationary (for example, the gross
national product time series \cite{granger}).
It is shown in \cite{peng2} that after the trend is
removed, the $var_y(l)$ is a linear function of $l$.
It is also shown in \cite{peng2} by studying some controlled sequences
that after removing the trends, the $\log(var_y(l))$ versus
$\log(l)$ can have either a non-trivial scaling relation
or a trivial linear relation. In other words, G+C content is
not responsible for all features of the correlation structure in DNA sequences.

\item
The suggestion that correlation will disappear when
one sample statistic from a sub-region with stable G+C content
requires that the boundary of such sub-region can be easily
delineated. As can be seen in Figure 1, it is not at all
easy. Although we can say (see Figure 1(e)) that there
are two G+C rich sub-regions in budding yeast chromosome 3,
if we look at the finer scaled base composition plot such
as Figure 1(b) or (c), there are also G+C poor regions
within the supposedly G+C rich region. There will still
be correlations if we only sample statistics from 
one of the two G+C rich regions as delineated by Figure 1(e).

\item
The whole issue comes down to the question of whether a DNA sequence
is simple or complex. For simple sequences, there are only 1 or
2 length scales, and a feature of the correlation structure can be
easily traced to its source and thus explained.
For complex sequences, there are many different length scales,
and one may not even be able to spatially localize the contribution
of a particular length scale. 

\end{itemize}

Actually, the correlation structure of sequences that are
decomposable to sub-regions (and each one exhibits white noise)
can easily be calculated. Figure 6 is a schematic drawing
of the situation: (a) for two white-noise sub-regions, and
(b) for five such regions. The detailed calculation of
the correlation structure (covariance function) is carried out
in Appendix.

In summaries, if the base composition of
the two sub-regions are different, the overall covariance
function is approximately equal to  a weighted sum of the
covariance functions from each sub-region plus a constant term contributed
from the base density differences between each sub-region
pair. The extra term from the base density difference
is always positive.

In the derivation (see Appendix), we assume the distance $d$
is smaller than any sub-region size. As a result, the
base-pairs are mostly taken from within a sub-region
rather than from between two sub-regions. The cross-subregion
sampling of base-pairs can contribute a $d$-dependent term
to the covariance function, but such contribution is
proportional to $d/N$ (distance divided by the sequence
length) which is very small. If $d$ is much larger than
all sub-region sizes, the concept of sub-region ceases
being relevant because at this distance $d$, the
overall correlation doesn't depend on the correlation
structure for each individual sub-region.

\vspace{0.4in}

{\bf On the difference of correlation structure between
coding and non-coding regions}:

In higher organisms, the mapping of a stretch of DNA to the corresponding
amino acid sequence is frequently  not continuous, but is interrupted by the
regions called ``introns". Introns
are non-coding regions, since its DNA contents do not translate to
part of the amino acid sequence. The parts that are translated
to the amino acid sequence are called ``exons". Much effort is  spent
on automatic recognition of intron/exon regions, and currently the
success rate for a computer recognition is about 60 -- 70 \%  \cite{fick-tung}.

If we know which sequence is coding and which is non-coding,
we can determine the correlation structure for each sequence, and
then see whether there is a common property among all coding
sequences or among all non-coding sequences. There does not seem
to be {\sl a priori} reason that coding sequences should have a
different correlation structure from that of non-coding sequences.

In \cite{fick-torney}, the correlation coefficient between the
base composition in two co-moving windows is studied as a function of the
distance between the two windows. This correlation coefficient
decays much slower in human DNA than {\sl E. coli} DNA \cite{fick-torney}.
Non-coding regions are common, of course, in human DNA, and
rare in {\sl E. coli}.

In \cite{wli-dna}, it is observed that non-coding sequences
consistently have longer correlation length than coding
sequences. Nevertheless, it is not clear whether the result
can be generalized to other cases because there are few sequences
being examined and all sequences are human DNA.

In \cite{peng}, it is shown that all complementary DNA (cDNA)
(they are made by reverse transcription from the mature mRNA
and thus do not contain introns) and other
intron-less DNAs such as prokaryote DNAs  being studied do not exhibit 
long-range correlation (judged by whether $var_y(l)$ is a linear function of $l$).
However many intron-containing DNAs do have long-range correlation
(judged by whether $var_y(l)$ is a non-linear function of $l$).
Two more examples are given in \cite{buldyrev-prl} showing again that
intron-less sequences do not exhibit long-range correlation.

One suggestion is that intron-containing sequences have
long-range correlation because intron and exons may
have totally different statistical properties and when they
are mixed in the sample statistics, a spatial structure
might be detected \cite{nee}, as shown for example by
the calculation in Appendix (e.g., Eq.(A.\ref{eqn:multi}) ).
It implies that if the similar
calculation is carried out for intron only, no
long-range correlation will be detected. Such calculation
has already been done in \cite{wli-dna} and clearly intron 
alone can exhibit long-range correlation.

Instead of separating exon and intron, in Ref. \cite{voss}, DNA
sequences were grouped according to their Genbank categories.\footnote{
As pointed out in \cite{chat}, these categories are
not of the equal taxonomic rank.}
Besides the fact that
the exponent $\alpha$ in the $1/f^\alpha$ spectral component
is slightly different from one category to another \cite{voss},
there is a striking qualitative difference between the
spectra of bacteria and phage sequences and those of others,
Note that bacteria and phage sequences contain mostly coding sequences
--- these sequences are more ``compact" or ``efficient" in the
context of protein-making.

In another study \cite{voss-prl}, it is shown that
in a certain grouping among a few GenBank categories
(group 1 contains primate, rodent, mammal, vertebrate DNAs,
group 2 contains invertebrate, plant DNAs, and group 3
contains virus, organelle, phage DNAs), intron sequences
and exon sequences do not exhibit dramatic difference
in their $1/f^\alpha$ spectral component. Some of the
grouping is questionable, such as whether organelle
DNAs share anything in common with virus DNAs.
Also, one important question not addressed is about
non-coding sequences
other than introns, i.e., the intergenic sequences: whether intergenic
sequences have a different correlation structure from that of intron
sequences. 

Despite the inconclusiveness of our knowledge concerning the extent
the correlation structure differs between coding and non-coding 
sequences, we make the following speculations:
\begin{itemize}
\item
As mentioned before, there is no {\sl a priori} reason
to believe that the correlation structure should be different
between coding and non-coding sequences. However, there are
major physical characteristics of coding sequences, most
notably the constraint of triplet (codon) usage in translation
(for a review of this topic see several chapters in \cite{doolittle-book}).  
One reason to expect  that correlation structure might be different
in the two is that coding regions are likely to be under different 
evolutionary constraints than non-coding regions.

\item
If this is correct, the way correlation
structure diverges between coding and non-coding sequences
would depend on what have been the main driving force in the DNA
changes in non-coding regions, and how often each mechanism
of DNA change has been in effect. For example, if
point mutation is the major driving force to change non-coding
regions, non-coding regions would look more like a random sequence than
coding regions. On the other hand, if duplication of larger
segments is the major driving force, non-coding regions would become
less random and more regular than coding regions.

\item
Will the difference of correlation structure between coding and
non-coding regions ever be useful in an algorithm for automatic
recognition of coding regions? It would be
less useful if other methods are more accurate and if
the DNA sequences are known accurately. But if the sequence
data is error-prone (it might be the case for a brute-force
sequencing in the early stage of human genome project),
one needs to re-evaluate each method \cite{states}.  Since
the calculation of correlation structure is rather insensitive
to a frame shift due to point deletion or point insertion, the result
will be the same whether such error exists or not in
the sequence.

\end{itemize}

Finally, since it has been shown that long-range correlation 
exists of the DNA sequence from budding yeast chromosome 3 \cite{mzhang,munson},
it is interesting to note that 30\% of the sequence is estimated to be  non-coding
and most of them are intergenic sequences instead of introns \cite{guigo}.

\section{Dynamical Origins of Long-range Correlation in DNA Sequences}

If we can recreate the dynamical process that led to the current
DNA sequences, we should be able to understand why these sequences
have the correlation structure they have today. For example,
we can take the point of view that the Markov chain model
implicitly assumes that the sequence is
created from one end to another: at each moment, a new site
is considered, and a new base is emitted from the source of the Markov
chain to become the last base of the sequence. The probability of which 
type of base is picked depends on the previous base (or the previous few bases
for higher-order Markov chains)
in the sequence. Such a process is not clearly a natural one for
modeling DNA evolution, and it is thus not surprising that the
Markov chain fails to characterize many features of the
correlation structure of DNA sequences.

A different class of models carries out sequence manipulation,
i.e., it updates an existing sequence using certain rules. For
example, cellular automata are a type of sequence manipulation
that synchronously updates bases according to the local environment
only (see \cite{ca1,ca2} for a general discussion and \cite{hag}
for the recent work). The idea to use cellular automata
to model DNA sequence change is considered in Ref.\cite{burks-farmer}.
We might ask the following question: what is the resulting correlation structure 
if one applies a cellular automaton rule repeatedly on a sequence
that is initially  white noise?

This question is studied in \cite{wli87}. The answer is provided only
for a special class of situations:   if the dynamics of
a cellular automaton is periodic, the sequence can then be
characterized by regular languages (almost the same as
Markov chain), and the autocorrelation function of the
limiting sequence is exponential --- long-range correlation
occurs (if it does) only within the framework of exponential
functions \cite{wli87}.  For more general situations, 
we make a few comments:
\begin{itemize}
\item
Since cellular automata updates bases locally, it is usually hard
to propagate local effects cooperatively to generate a long-range
correlation.

\item
For some cellular automata, their temporal dynamics are irregular and
their spatial configuration is unstable -- these are termed ``chaotic".\footnote{
To avoid confusion, this definition for a cellular automaton being
chaotic \cite{norm,caspace} is different from the definition 
in continuous-variable non-linear dynamical systems:
here, we have {\sl linear} propagation of perturbation in
{\sl position space}, there, it is the {\sl exponential} propagation
of perturbation in {\sl variable space}. One way to reconcile
the two is to map a (e.g. binary) sequence to a real value
so that the binary representation of the real value is that
binary sequence with a particular choice of the decimal point.
By doing so, a linear divergence in the  position space
corresponds to an exponential divergence in the variable space.
}
Since a local effect propagates quickly in chaotic cellular
automata, we would expect some long-range correlation exists
in the limiting sequence. Nevertheless, such long-range correlation
tends to be a weak signal because there is a high level of randomness
created by the chaotic dynamics. In this case, the long-range correlation
is expected to be statistically insignificant.

\item
The best candidate in cellular automata for generating detectable
long-range correlation has a combination of the ability to
propagate local effects and a low level of randomness. Such
rules are called ``edge-of-chaos" cellular automata \cite{langton,caspace}.
In fact, because the correlation at intermediate ranges is
small for cellular automata with either periodic or chaotic
dynamics, the existence of such correlation is able to locate
the edge-of-chaos region in a cellular automata rule space \cite{caspace}.

\item
In a special class of cellular automata that maintain propagating
gliders (solitons), the spatial pattern can be very complicated \cite{aizawa}.
In several cases, the spatial spectra after averaged over 500 time steps
are $1/f^\alpha$ ($\alpha > 0.5$) at low frequencies (see Figure 4.2(b),
4.7, and 4.9 of \cite{aizawa}). But note that such low-frequency
$1/f^\alpha$ spectra often appear during the transient, also
note that different length scales may be picked up at different
time steps (because a time average has been performed).

\end{itemize}

Since cellular automata are again not clearly the most ``natural" model
for DNA evolution, we turn our attention to sequence manipulations
that increase the sequence length. In the following three
sub-sections, we will first review the ``expansion-modification"
system, and then argue about the relevance of models like
expansion-modification systems to DNA evolution.
Some of our point is presented in \cite{wli-kuni-2}.

\subsection{Expansion-modification systems}

The prototype of the expansion-modification systems is
defined as follows (for two symbols) \cite{wli89,wli91}:
\begin{eqnarray}
\label{eqn:exp-mod}
        t & & t+1 \nonumber  \\
        1 &\rightarrow& \left\{ \begin{array}{lcl}
                11 &:& \mbox{$1-p$} \\
                0 &:& \mbox{$p$}
                \end{array} \right. \nonumber \\
        0 &\rightarrow& \left\{ \begin{array}{lcl}
                00 &:& \mbox{$1-p$} \\
                1 &:& \mbox{$p$}.
                \end{array} \right.
\end{eqnarray}
To describe in words, a symbol 1 at time $t$ is updated to either two
symbols 11 (with a probability $1-p$), or a symbol 0 (with a
probability $p$). Similar action is applied to  symbol 0.
When the probability for the ``switch" operation ($p$) is
small, the limiting sequence exhibit a perfect $1/f^\alpha$ 
($\alpha \approx 1$) spectrum (unlike DNA sequences with
$1/f^\alpha$ spectral component, spectra of sequences
generated by Eq.(3.\ref{eqn:exp-mod}) do not contain a white
noise component). The exponent $\alpha$ is a function of $p$.

Unlike cellular automata, the expansion-modification system uses
a different way to propagate local effects to global scale:   it
forces the old neighbors farther away by creating new neighbors. 
This mechanism avoids the difficulty of having both elements of
chaos and non-chaos at the same time as for the case of a cellular automaton.

A simple argument can show that power-law covariance 
or autocorrelation functions can be {\sl maintained} by
the expansion-modification system. The argument goes as follows
(we use subscript to represent the time step): first, there
is a linear expansion of the distance after each time step:
\begin{equation}
\label{eq:distance}
d_{t+1} \approx d_t (2-p). 
\end{equation}
Then, the covariance at time step $t$, distance $d_t$
is essentially the same with that at time step $t+1$, distance $d_{t+1}$,
with perhaps a proportionality factor $\lambda$.  Using Eq.(\ref{eq:distance}),
we have
\begin{equation}
cov(d_{t+1})_{t+1} \approx \lambda cov(d_t)_t
\approx  \lambda cov( \frac{d_{t+1}}{2-p})_t.
\end{equation}
Assuming that the covariance function is invariant by the
updating (another way to say this is that the function reaches the
asymptotic limit), so we can drop all subscripts:
\begin{equation}
\label{eq:inv}
cov(d) \approx \lambda cov(\frac{d}{2-p}).
\end{equation}
A power-law function $cov(d) \sim 1/d^\beta$ is a solution
of Eq.(\ref{eq:inv}), as long as
\begin{equation}
 \beta = - \frac{log(\lambda)}{ log(2-p)}.
\end{equation}
When $\beta \approx 0$, the power spectrum $1/f^{1-\beta}$ becomes $1/f$.
For $\beta$ being 0 requires $\lambda$ being close to 1, but $p$ away from 1.

This argument is used in \cite{wli89,wli91}.  Actually, the condition
for $cov(d)$ being 1-step invariant can be relaxed to $\tau$-step
($\tau >1$) invariant:
\begin{equation}
cov(d) = \lambda^\tau cov( \frac{d}{ (2-p)^\tau})
\end{equation}
Since we only deal with positive, real values here, as
the function $cov(d) \sim 1/d^\beta$ is inserted to the equation,
the expression for $\beta$ is the same. As $\tau$
goes to infinity, and the same result holds.

Note that our argument about power-law covariance function
being maintained by some rule with a linear expansion of distance
does not really show {\sl how} the power-law covariance function
is {\sl generated}, {\sl  before being maintained}. The latter might be
called an ``operational" proof.

\subsection{Duplication Events in DNA Evolution}

We use the phrase "DNA duplication" in a general way here to describe the
general situation when segments of DNA get copied and inserted elsewhere in
the genome. This can include situations when DNA is duplicated to adjacent
segments, or otherwise. The result in either situation is the development
of tandem (adjacent) or dispersed repeating segments. Because of the underlying
mechanisms involved, the duplicated segments can appear as not exact copies.

The expansion-modification system described in Eq.(3.\ref{eqn:exp-mod})
represents one special
case of duplication, that involves single point duplication that leads to a
single tandem repeat per duplication event. Though this involves a simple 
situation, it is one that is thought to be common during DNA replication where
the replication machinery causes a single base to be duplicated through unequal
pairing during mitosis. However, this clearly represents a simple situation and
we make no claim about the biological reality of this model.  

Besides the duplication processes discussed above, there are many other sources
of variation generation in DNA sequences. They fall into two broad categories,
first there are the point mutations, such as deletions and insertions, and secondly
there are the chromosomal level variations such as recombination, translocations,
and deletions, for example. Chromosomal level (meaning essentially any relatively
large segment) variation processes are thought to be major sources for generation
of raw material for evolution to work on, such as in the shuffling of modular
elements described in \cite{doolittle}. 

The duplication of relatively large segments of DNA provides interesting sources
of variation for evolution. For example, entire genes are known to duplicate and
insert elsewhere in the genome, leaving the original copy intact and operational,
while the duplicated copy can be subject to further modification and evolutionary
selection. We will not address the level of organization at which evolution operates
here, but rather assume that a mechanism exists that can ultimately influence
the evolutionary potential of duplicated segments. Perhaps no one more advocates
the role of DNA duplication in evolution than S. Ohno \cite{ohno}. The principle 
argument here is that when entire genes or otherwise functional units are duplicated,
the working original is left alone, while the duplicated copy can be subject to
a different set of evolutionary dynamics.

While genome length per se does not scale with complexity of the organism, it is
true that longer genomes on an absolute scale can contain a larger collection of
distinct elements than smaller ones. Increased genome length puts additional
constraints on the existence of certain functional elements, such as origins of
replication, since replication during the cell cycle is time-constrained, so that
more origins are needed to replicate the longer segments during a fixed period
of time. There may be other similar constraints that exist that are necessary for
higher-order genome structure that are related to linear dimension. The type
of analysis we have described here may provide insight into what the dynamics and
constraints are in DNA organization. Clearly, there is much room for this type
of analysis as the international human genome program provides biology the opportunity
to ask such fundamental questions regarding DNA structure, organization, and dynamics.

\section*{Acknowledgements}

W.L. would like to thank Chung-Kang Peng, Michael Zhang, Roderic Guig\'{o}
for communicating results before publication, and Jim Fickett for providing
valuable comments to the first draft of the paper. The participation of W.L.
to the Oji International Symposium on Complex Systems was made possible
by a financial support from the Fujihara Foundation and Japanese Society
for Promotion of Science. The work of W.L. and T.M.  at Cold Spring
Harbor Laboratory is supported by the grant from DOE (DE-FG02-91ER61190).
T.M. is also supported by NIH grant 2-R01-HE0020304.

\newpage

\appendix
 
\begin{center}
APPENDIX
\end{center}
\section{Calculation of Covariance Functions for Sequences
Decomposable to Many White-noise Sub-regions}

\vspace{0.3in}
{\bf Two Sub-region Case (Same-Symbol Correlation)}:
Figure 6(a) illustrates this situation. The overall sequence length
is $N$, and the length for two sub-regions are $N_1$ and $N_2$.
The ratio of the two sub-regions relative to the whole is
$ f_1 \equiv N_1/N$ and $f_2 \equiv N_2/N$.
 
Suppose the count of nucleotide type $\alpha$ in two sub-regions
are $n_{\alpha(1)}$ and $n_{\alpha(2)}$ respectively, we have
\begin{equation}
\hat{P_\alpha} = \frac{n_\alpha}{N} = \frac{ n_{\alpha(1)} + n_{\alpha(2)}}{N}
= \frac{N_1}{N} \widehat{P_{\alpha(1)}} + \frac{N_2}{N} \widehat{P_{\alpha(2)}}
= f_1 \widehat{P_{\alpha(1)}} + f_2 \widehat{P_{\alpha(2)}}
\end{equation}
 
A similar decomposition of the joint probability is
\begin{equation}
\widehat{P_{\alpha\beta}(d)}=
\frac{n_{\alpha\beta(1)}(d)+n_{\alpha\beta(2)}(d) + n_{\alpha\beta(12)}(d) }{N}
= f_1 \widehat{P_{\alpha\beta(1)}(d)} +
f_2 \widehat{P_{\alpha\beta(2)}(d)} + \frac{n_{\alpha\beta(12)}(d)}{N}
\end{equation}
where $n_{\alpha\beta(12)}(d)$ is the number of cross-subregion counts
of the ($\alpha$, $\beta$) base-pair out of total $d$ counts.  It is
usually negligible as compared with other two terms if $d << N_1, N_2, N$.
 
The differences of base composition between two sub-regions are:
\begin{equation}
\Delta P_\alpha \equiv P_{\alpha(1)} - P_{\alpha(2)} \ne 0
\hspace{0.3in}
\Delta P_\beta \equiv P_{\beta(1)} - P_{\beta(2)} \ne 0.
\end{equation}
It can be shown that the covariance function of the whole sequence
as defined by Eq.(\ref{eq:est-cov}) is related to the covariance functions
of the two sub-regions in the following way:
\begin{eqnarray}
\label{eqn:same-base}
\widehat{cov(d)} &=& \sum_{\alpha =(G,C,T,A)} 
\left( \widehat{P_{\alpha\alpha}(d)} - \widehat{P_\alpha}^2
\right) \nonumber\\
&=& f_1 \widehat{cov_1(d)} + f_2 \widehat{cov_2(d)}
+ f_1 f_2  \sum_{\alpha = (G,C,T,A)} ( \widehat{\Delta P_\alpha} )^2
+ \sum_{\alpha = (G,C,T,A)} \frac{n_{\alpha\alpha(12)}(d)}{N}
\nonumber \\
&\approx&  f_1 \widehat{cov_1(d)} + f_2 \widehat{cov_2(d)}
+ f_1 f_2  \sum_{\alpha = (G,C,T,A)} ( \widehat{\Delta P_\alpha} )^2
\hspace{0.3in} \mbox{(if $d << N_1, N_2, N$)} \nonumber \\
&\approx& f_1 f_2 \sum_{\alpha = (G,C,T,A)} ( \widehat{\Delta P_\alpha} )^2
\hspace{0.3in} \mbox{(if sub-regions are white noise and $d>0$)}
\end{eqnarray}
The above formula shows that the difference of base composition alone
contributes a {\sl constant} term to the overall covariance or
autocorrelation function. The  $d$-dependent term of $cov(d)$
is of the order of $o(d/N)$ if $d$ is smaller than $N_1, N_2$:
\begin{equation}
\sum_{\alpha = (G,C,T,A)} \frac{n_{\alpha\alpha(12)}(d)}{N}
\approx \sum_{\alpha = (G,C,T,A)}
\frac{d}{N} \widehat{ P_{\alpha(1)}} \widehat{ P_{\alpha(2)}}.
\end{equation}

To summarize, the overall correlation structure of the sequence,
if it is decomposable to two sub-regions with white noise but
different base compositions, is very simple.  It does not share
the similar feature with complex sequences of the multi-length-scales
and $1/f^\alpha$ spectral component.
 
\vspace{0.3in}

{\bf Two Sub-region Case (Different-Symbol Correlation)}: One can
also easily derive the correlation between bases
of different type ($\alpha \ne \beta$):
\begin{eqnarray}
\label{eqn:diff-base}
\widehat{cov_{\alpha\beta}(d)}
&=& \widehat{P_{\alpha\beta}(d)} - \widehat{P_\alpha} \widehat{P_\beta} \nonumber \\
&\approx& f_1  \widehat{cov_{\alpha\beta(1)}(d)}
+ f_2  \widehat{cov_{\alpha\beta(2)}(d)}
+ f_1 f_2 \widehat{ \Delta P_\alpha} \widehat{ \Delta P_\beta}
\hspace{0.3in} \mbox{(if $d << N_1, N_2, N$)} \nonumber \\
&\approx& f_1 f_2 \widehat{ \Delta P_\alpha} \widehat{ \Delta P_\beta}
\hspace{0.3in} \mbox{(if sub-regions are white noise and $d>0$)} \nonumber \\
&=&  - f_1 f_2 (\widehat{\Delta P_\alpha})^2
\hspace{0.3in} \mbox{(if the sequence is binary)}
\end{eqnarray}
Interestingly, the observation in the budding yeast chromosome 3
sequence  that $\widehat{I_{\alpha\alpha}(d)}$'s tend to be larger than 1,
while $\widehat{I_{\alpha\beta}(d)}$'s ($\beta \ne \alpha$)
tend to be smaller than 1 \cite{mzhang} can be compared with
our result Eq.(A.\ref{eqn:same-base}) and Eq.(A.\ref{eqn:diff-base})
that covariance functions between same base type acquire a positive
contribution from the base composition difference, whereas
those between different base types acquire a negative contribution.
 
\vspace{0.3in}
{\bf Many Sub-regions}
All the above results can be generalized to $K>2$ different
sub-regions, each of them is a white noise. Figure 6(b) illustrates
the situation, with the overall sequence length as $N$, the
length of $i$'th sub-region as $N_i$, and $f_i \equiv N_i/N$.
For example, we can show that
\begin{eqnarray}
\label{eqn:multi}
\widehat{cov(d)} &=&
\sum_{\alpha= (G,C,T,A)} \left[
\sum_{i=1}^K f_i \widehat{P_{\alpha\alpha(i)}(d)} +
 \sum_{i=1}^{K-1} \frac{n_{\alpha\alpha (i,i+1)}(d)}{N}
- \left(\sum_{i=1}^K f_i \widehat{P_{\alpha(i)}} \right)
 \left(\sum_{j=1}^K f_j \widehat{ P_{\alpha(j)}} \right)
\right]
\nonumber \\
&=& \sum_{i=1}^K f_i \widehat{ cov_i(d) } +
\sum_{\alpha= (G,C,T,A)} \sum_{i=1}^K \sum_{j>i}
f_i f_j  (\widehat{ \Delta P_{\alpha(ij)}} )^2 +
\sum_{\alpha= (G,C,T,A)} \sum_{i=1}^{K-1} \frac{n_{\alpha\alpha (i,i+1)}(d)}{N}
\nonumber \\
&\approx& \sum_{i=1}^K f_i \widehat{ cov_i(d) } +
\sum_{\alpha= (G,C,T,A)} \sum_{i=1}^K \sum_{j>i}
f_i f_j  (\widehat{ \Delta P_{\alpha(ij)}} )^2
\hspace{0.3in} \mbox{(if $d << N_i, N$)} \nonumber \\
&\approx&
\sum_{\alpha= (G,C,T,A)} \sum_{i=1}^K \sum_{j>i}
f_i f_j  (\widehat{ \Delta P_{\alpha(ij)}} )^2
\hspace{0.1in} \mbox{(if sub-regions are white noise and $d>0$)}
\nonumber \\
\end{eqnarray}
where $n_{\alpha\alpha(i,i+1)}$ is the count of ($\alpha, \alpha$)
base-pairs crossing the $i$'th and the $(i+1)$'th sub-regions.


\newpage

\begin{figure}
\caption{
\label{fig:1}
$G+C$ content and purine density of budding yeast chromosome 3
with various window sizes ($W$) and shifting distances ($D$).
The sequence length is $N=315,338$.

(a) $W=1000$ and $D=100$;

(b) $W=1000$ and $D=1000$: non-overlapping windows;

(c) $W=10,000$ and $D=1000$;

(d) $W=20,000$ and $D=1000$; and

(e) $W=100,000$ and $D=2000$.
}
\end{figure}

\begin{figure}
\caption{
\label{fig:2}
$G+C$ content and purine density of  human cytomegalovirus (strain AD169)
with various window sizes ($W$) and shifting distances ($D$). The sequence
length is $N=229,354$.

(a) $W=1000$ and $D=1000$: non-overlapping windows;

(b) $W=5000$ and $D=1000$; and

(c) $W=10,000$ and $D=1000$.
}
\end{figure}

\begin{figure}
\caption{
\label{fig:3}
$G+C$ content and purine density of  human blood coagulation factor VII gene
with various window sizes ($W$) and shifting distances ($D$). The sequence
length is $N=12,850$.

(a) $W=30$ and $D=30$: non-overlapping windows;

(b) $W=100$ and $D=30$; and

(c) $W=1000$ and $D=30$.
}
\end{figure}

\begin{figure}
\caption{
\label{fig:4}
$G+C$ content and purine density of bacteriophage lambda
with various window sizes ($W$) and shifting distances ($D$). The sequence
length is $N=48,502$.

(a) $W=100$ and $D=100$: non-overlapping windows;

(b) $W=2000$ and $D=100$; and

(c) $W=10,000$ and $D=100$.
}
\end{figure}

\begin{figure}
\caption{
\label{fig:5}
Superposition of Lorenzian-like spectra (Eq.(\ref{eq:loren2})).
Sequence length is $N=50,000$ and frequencies are $f = i/N$ ($i=1, 2, \cdots N/2$):

(a) case (1) is a single Lorenzian spectrum:  $a(d_0=200)=0.03$;
case (2) contains two Lorenzian spectra: $a(d_0=200)=0.03$ and $a(d_0=500)=0.01$;
case (3) contains  three Lorenzian spectra: $a(d_0=200)=0.03$, $a(d_0=500)=0.01$
and $a(d_0=2000)=0.007$.

(b) Same with (a) except that the sequence length is longer: $N=500,000$.

(c) Similar to (a) but $a(d_0)$'s are normalized to 1: (1) $a(d_0=200)=1$;
(2) $a(d_0=200)=0.75$, $a(d_0=500)=0.25$; and
(3) $a(d_0=200)=0.6383$, $a(d_0=500)=0.2128$, and $a(d_0=2000)=0.1489$.
}
\end{figure}

\begin{figure}

\vspace{1.0in}
 
\thicklines
\begin{picture}(200,80)(0,50)
 

\put (30,150){\framebox(140,30)}
\put (170,150){\dashbox{5}(160,30)}

\put (390, 165) {(a)}
\put (100,135) { $N_1$ }
\put (240,135) { $N_2$ }
\put (150,195) {$N=N_1+N_2$}

\put (30,50){\framebox(60,30)}
\put (90,50){\dashbox{5}(35,30)}
\put (125,50){\framebox(75,30)}
\put (200,50){\dashbox{5}(80,30)}
\put (280,50){\framebox(50,30)}

\put (390, 65) {(b)}
\put (50,35) { $N_1$ }
\put (94,35) { $N_2$ }
\put (150,35) { $N_3$ }
\put (225,35) { $N_4$ }
\put (295,35) { $N_5$ }
\put (160,94) {$N=\sum_i N_i$}

\end{picture}
\vspace{1.3in}

\caption{
\label{fig:6}
Illustration of the situation when the sequence can be decomposed
to (a) two sub-regions, or (b) many sub-regions, with each sub-region
being a white noise, but different base compositions.
}
\end{figure}

\end{document}